# Band splitting and Modal Dispersion induced by Symmetry braking in Coupled-Resonator Slow-Light Waveguide Structures

Jacob Scheuer[1] and Mark Shtaif
*School of Electrical Engineering, Tel-Aviv University, Ramat-Aviv, Israel*
[1]kobys@eng.tau.ac.il

**Abstract**: We study the dispersion relations in slow-light waveguide structures consisting of coupled microdisk resonators. A group theoretical analysis of the symmetry properties of the propagating modes reveals an interesting phenomenon: The degeneracy of the CW and CCW rotating modes is removed, giving rise to two distinct transmission bands. This effect induces symmetry-based dispersion which may limit usable bandwidth of such structures. The properties of this band splitting and its impact on CROW performance for optical communications are studied in detail.
PACS numbers:

## I. INTRODUCTION

The symmetry properties of any physical system is one of the most profound and basic mechanisms underlying the system physics and dynamics. Symmetry and, especially, symmetry braking phenomena provide crucial insight and determine many of the basic properties of systems in diverse fields of physics ranging from classical and quantum mechanics to solid-state physics and photonics [1]. In solid-state physics, for example, symmetry considerations and group theory have been used extensively to study the electronic properties and energy bands of crystals and semiconductors [2]. In photonics, symmetry has been used to study the bandgaps and passbands of Photonic Crystals (PhC) as well as the confined eigen-modes of PhC defect cavities [3]. The role of symmetry in circular cavities has also been studied and various applications have been proposed [4]

In the past few years, much attention was devoted to slowing down the propagation speed of light and to coherently stop and store pulses of light [5-8]. In particular, significant efforts were focused on controlling the speed of light using photonic structure incorporating microcavities. Substantial delays and storage of light pulses were predicted in various coupled-cavities structures such as coupled resonator optical waveguides (CROWs) [9] and side-coupled integrated spaced sequence of resonators (SCISSORs) [10]. The ability to control the speed of light in chip-scale components and realize ultra-compact optical delay lines is highly desired and much effort was focused on the reduction of micro-resonator dimensions and on increasing their density.

Microring and microdisk resonators based slow-light structures (SLS) draw much attention for practical applications, primarily because they are relatively simple to fabricate and characterize. In contrast, the fabrication and characterization of PhC defect cavities based structures is more challenging and necessitate sophisticated high resolution fabrication tools as well as special arrangement for I/O coupling.

One of the important properties of microdisk resonators is the inherent degeneracy of clockwise (CW) and counterclockwise (CCW) whispering gallery modes (WGM). In the analysis of linear CROW and SCISSOR structures, this degeneracy is generally disregarded because it is implicitly assumed that its only impact is doubling the dispersion relation [10, 11]. As a results, the general properties of microdisk resonator and PhC defect cavities based CROWs are considered to be very similar. Nevertheless, it has been shown that in some scenarios this degeneracy results in profound differences between the properties of PhC cavities (i.e. non-degenerate) and micro-ring/disk based structures. An example for such scenario is the employment of coupled resonator structures for rotation sensing [12-14].

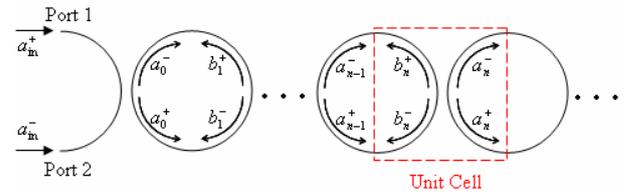

Figure 1. Schematic of a CROW structure and field amplitudes definitions.

In this paper, we analyze the properties of the propagating modes in microdisk based CROWs (see Fig. 1). We employ the tools of group theory and find that, unlike a single microdisk, CROW structures *do not* possess band degeneracy and that their Bloch modes *always* comprise standing-wave patterns in the microdisks. The impact of this degeneracy removal is that for each frequency there are four distinct Bloch wavenumbers (two in each propagation direction). In other words, a CROW structure exhibits two distinct transmission bands corresponding to symmetric and anti-symmetric superposition of the CW and CCW WGM in each unit cell. Because CROW structures are generally excited by injecting light into one of the input ports (e.g. port 1 – see Fig. 1) this existence on the non-degenerate bands induces additional dispersion which could be substantial especially for wide bandwidth applications.

## II. TRANSFER MATRICES AND SYMMETRY PROPERTIES

The electric field in a CROW can be represented by a vector of four amplitudes of the CW and CCW

propagating waves in the two half-rings constituting the unit cell (see fig. 1).

$$v_n = \begin{bmatrix} a^+ & a^- & b^+ & b^- \end{bmatrix}_n^T \quad (1)$$

In the linear optics regime, the fields vector in the $n$th unit cell can be related to that in the $n+1$ cell by a transfer matrix [10]:

$$v_{n+1} = \mathbf{M} \cdot v_n \quad (2)$$

where $\mathbf{M}$ is a 4x4 matrix. Because of the periodicity of the CROW structure, the propagating modes must also satisfy the Floquet-Bloch theorem which requires the amplitudes-vector in each unit cell is identical except for a phase difference:

$$v_{n+1} = v_n \exp(-iK\Lambda) \quad (3)$$

which means that a Bloch mode of the structure must satisfy the eigen-value problem:

$$(\mathbf{M} - e^{-iK\Lambda}) \cdot v = 0 \quad (4)$$

where $K$ is the Bloch wavenumber and $\Lambda$ is the periodicity of the CROW. Because of the geometrical symmetry of the structure, one may apply appropriate rotation and reflection operators on the structure and the propagating modes (according to the group symmetry) resulting in a legitimate, but different propagating mode. Denoting such operator as $\mathbf{D}_i$ then for any solution of (4) there are additional solutions satisfying:

$$(\mathbf{M}' - e^{-iK\Lambda}) \cdot v' = 0 \quad (5)$$

where $v' = \mathbf{D}_i \cdot v$ and $\mathbf{M}' = \mathbf{D}_i \cdot \mathbf{M} \cdot (\mathbf{D}_i)^{-1}$. In principle, (5) holds for any invertible transformation $\mathbf{D}_i$. However, under a group symmetry operation, the structure is transformed back to itself and thus, the transformed matrix $\mathbf{M}'$ is necessarily identical to the original one $\mathbf{M}$. Equivalently, this equality means that the transfer matrix $\mathbf{M}$ must commute with all the group symmetry operations:

$$[\mathbf{M}, \mathbf{D}_i] = 0 \; \forall \; \mathbf{D}_i \in g \quad (6)$$

where $g$ denotes the pertinent symmetry group.

### III. GROUP THEORETICAL ANALYSIS

For the CROW structure illustrated in Fig. 1, the group symmetry is $C_{2V}$ with the covering operations $\mathbf{E}$, $\mathbf{C_2}$, $\mathbf{\sigma_x}$, $\mathbf{\sigma_y}$. $\mathbf{E}$ is the identity transformation, $\mathbf{C_2}$ is a 180° rotation around the vertical axis, and $\mathbf{\sigma_x}$, $\mathbf{\sigma_y}$ are reflection operations around the $x$ and $y$ axes respectively [15]. Appendix A summarizes the properties of the $C_{2V}$ group symmetry and lists the 4x4 matrix representation, $R$, of the covering operations. It is important to note that each operation in the $C_{2V}$ group forms a *class of its own* (table A1). This property is important as it has substantial impact on the number of appearances of each irreducible representation (and hence on its degeneracy) in our 4x4 representation.

The traces (characters) of the covering operation represented by $R$ are given in Table 1:

| E | $C_2$ | $\sigma_x$ | $\sigma_y$ |
|---|---|---|---|
| 4 | 0 | 0 | 0 |

Table 1 – Traces of the $C_{2v}$ covering operation

Using the $C_{2V}$ character table and applying the appearance theorem [15] we find the number of times each representation appears in the 4x4 matrix representation:

$$\eta_\mu = \sum_\beta \frac{\Gamma_\beta}{\Gamma} \bar{\chi}^\mu(\beta) \chi^R(\beta) \quad (7)$$

where the sum is over all classes $\beta$ in $C_{2v}$, $\Gamma_\beta$ is the number of operations in class $\beta$, $\Gamma$ is the total number of operations, $\chi^\mu$ is the character of a group operation belonging to class $\beta$ represented by the irreducible representation $m$ and $\chi^R$ is the trace of the same operation represented by $R$. Introducing the parameters into (7) we find that $\eta_1 = \eta_2 = \eta_3 = \eta_4 = 1$, i.e., each representation appears only *once*. This is an important result with profound implications – it means that for each Bloch vector $K$ there are four distinct eigen-vectors (and frequencies) which are by no means necessarily degenerate. The four modes can be divided into two groups – right propagating modes and left propagating modes (two modes in each group) which for a given $K$ posses different frequencies. However, despite what may be expected due to the whispering gallery modes (WGM) degeneracy in each unit cell, the two co-propagating modes are not necessarily degenerate (see also section IV) and the intuitive picture of equivalent CW and CCW WGM based Bloch waves does not hold.

Next, we set to determine the Bloch wave $v_\mu$ corresponding to each representation. To achieve that, we need to find the projector operator of $R$ on each of the irreducible representation of $C_{2v}$. The projector of $R$ on the irreducible representation $\mu$ is given by [15]:

$$P^{\mu \leftarrow R} = \sum_g \frac{d_\mu}{\Gamma} \bar{\chi}^\mu(g) D_i(g) \quad (8)$$

where $d_\mu$ is the dimension of the irreducible representation $\mu$ and the sum is over all the operation in group $g$. $\chi^\mu(g)$ is the character of any of the group operations in representation $\mu$ and is given by the group character table (see table A1).

$$\begin{aligned} P^{A_1 \leftarrow R} &= \tfrac{1}{4}(E + C_2 + \sigma_x + \sigma_y) \\ P^{A_2 \leftarrow R} &= \tfrac{1}{4}(E + C_2 - \sigma_x - \sigma_y) \\ P^{B_1 \leftarrow R} &= \tfrac{1}{4}(E - C_2 + \sigma_x - \sigma_y) \\ P^{B_2 \leftarrow R} &= \tfrac{1}{4}(E - C_2 - \sigma_x + \sigma_y) \end{aligned} \quad (9)$$

And the eigen-vectors (i.e. Bloch waves represented according to the basis $v$), found by projecting $R$ on the irreducible representations:

$$v_1 = \begin{pmatrix} 1 \\ 1 \\ 1 \\ 1 \end{pmatrix}, \quad v_2 = \begin{pmatrix} 1 \\ -1 \\ 1 \\ -1 \end{pmatrix}, \quad v_3 = \begin{pmatrix} 1 \\ 1 \\ -1 \\ -1 \end{pmatrix}, \quad v_4 = \begin{pmatrix} 1 \\ -1 \\ -1 \\ 1 \end{pmatrix} \quad (10)$$

We find that the eigen-modes of the CROW structure are essentially standing waves. This should not come as a surprise. According to Floquet-Bloch theorem, a Bloch mode of a periodic structure is a product of a phase function by a periodic function with the *periodicity of the structure*. This means that the amplitudes vector, $v_n$ must be identical in all resonators comprising the structure. The WGM like solutions, on the other hand, *do not* possess such property because the fields in adjacent cavities circulate in *opposite* directions. Effectively, it means that the periodicity of the WGM-like solutions is *twice* that of the structure, i.e., the "unit-cell" must comprise two adjacent resonators. This property of the CROW modes is also revealed by conventional transfer matrix analysis of the CROW [11] where the transfer matrix for a unit cell is composed from free propagation in the microdisks and the coupling matrix of the directional coupling. Nevertheless, such analysis reveals profound connections between the Bloch wavenumbers of the standing waves based Bloch modes. In particular, it was found that at any frequency, the corresponding pair of propagating (standing-wave based) Bloch modes have the same Bloch wavenumber except for a π phase shift (i.e. a minus sign). As a result, symmetric and asymmetric superposition of these modes yield the CW and CCW WGM solution with the same Bloch phase shift over a (double) unit-cell, which means that these modes are degenerate.

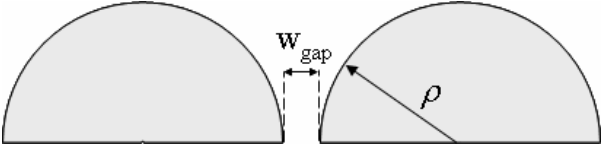

Figure 2. Dimensions and definitions of the simulated structure

However, the group theoretical analysis reveals an important property of the eigen-modes (Bloch waves) of a CROW: the irreducible representations appear only once and hence the eigen-modes are *non-degenerate*. Moreover, the eigen-modes *always* consist of either symmetric or anti-symmetric superposition of the CW and CCW rotating waves ($a^+ \pm b^-$, $b^+ \pm a^-$). We emphasize, that this outcome is found from a straightforward symmetry properties analysis of the CROW structure and, therefore, stems directly from the $C_{2v}$ group symmetry of the structure, regardless the specific shape of the resonators, coupling coefficients, etc.

The group symmetry analysis shows that there is no fundamental reason to assume degeneracy or inherent phase relations between the CROW Bloch modes. Therefore, injecting light into a CROW from a single input port (say, port 1 in Fig. 1) excites two Bloch modes exhibiting (in principle) *different* dispersion relations. The immediate result is additional (modal) dispersion which may limit substantially the bandwidth and delay that can be provided by the CROW.

It should be emphasized, that the degeneracy removal studied here does not stem from surface roughness and back scattering in the individual cavities [15, 16]. The analysis here assumed perfect microdisks and complete degeneracy of the CW and CCW propagating modes.

## IV. NUMERICAL SIMULATIONS

In order to verify and demonstrate the band splitting predicted in section III we calculated the dispersion relations of a coupled microdisks based CROW using the finite-difference-time-domain (FDTD) algorithm [18]. The simulated structure consists of a single unit cell where the Bloch boundary conditions are employed. The parameters of the studied structure are: disk radius $\rho = 0.7\mu m$, disk index $n_{disk}=3.5$, cladding index $n_{clad}=1.0$, separation $w_{gap}=0.5\mu m$ (see figure 2). The resonance wavelength of an individual resonator is $\lambda_{res}=1.565\mu m$ ($f_{res}=191.6THz$) with angular modal number $m=7$.

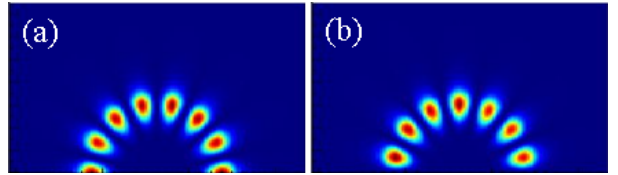

Figure 3. Symmetric (a) and anti-symmetric (b) modes of an individual microdisk.

Figure 3 depicts the symmetric and anti-symmetric modal field profiles of an individual microdisk as calculated by the FDTD simulations. Symmetric/Anti-Symmetric boundary conditions (BC) where employed in the lower part of the calculation window (-y) in order to separate between the two symmetries. Perfectly matched layers BC where used in the upper part of the calculation window (+y) as well as in the horizontal boundaries.

In order to calculate the dispersion relation of the CROW, Bloch boundary conditions were employed in the horizontal axis. The symmetric/anti-symmetric BC in the lower part were employed in order to separate between the symmetric and anti-symmetric branches of the dispersion relation. Note, that the simulated unit cell includes *two* disks. This, in order to account for the doubled periodicity of the WGM-based Bloch modes (if exists). Consequently, the simulations where conducted for Bloch wavenumber running from 0 to $2\pi$ in order to attain the complete dispersion relation.

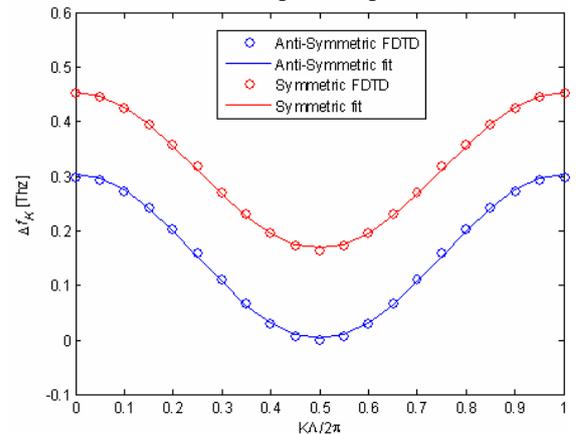

Figure 4. Dispersion relations of the Anti-symmetric (red) and symmetric (blue) bands of a CROW.

Figure 4 depicts the dispersion relations (DRs) of the propagating Bloch mode. An immediate result which is clearly seen is that the DRs of the symmetric and anti-symmetric brunches *differ significantly*. Figure 4 also shows a numerical fit of the FDTD results to a generic CROW dispersion relation obtained by the tight-binding approach [18]:

$$\Delta\omega_K = \tfrac{1}{2}\Omega\Delta\alpha - \kappa\Omega\cos(K\Lambda) \qquad (11)$$

where $\Delta\omega_K = \omega_K - \Omega$ is the difference between the optical frequency and the resonance frequency of an individual microring. $\kappa$ represent the coupling between the adjacent microdisks and $\tfrac{1}{2}\Delta\alpha\Omega$ is the self frequency shift [19]. Excellent agreement is found between the numerical results and analytical expression with the following parameters: $f\Delta\alpha_{sym} = 0.6236$ THz, $f\Delta\alpha_{a\text{-}sym} = 0.3072$ THz, $f\kappa_{sym} = 0.1426$ THz and $f\kappa_{a\text{-}sym} = 0.1497$ THz where $f$ is the resonance frequency of the cavity. It should be noted that while the coupling coefficients of the symmetric and anti-symmetric modes are quite similar, there is a non-negligible difference in the self frequency shifts of the two modes. It is this difference that accounts primarily for the deviation between the dispersion relations of the two branches. To verify that the observed splitting is not a numerical artifact, the simulations were run under various conditions and resolutions. No changes in the resonance frequencies were found for higher resolutions and longer integration periods. In addition, the resonance frequency of a *single* resonator was calculated separately for symmetric and anti-symmetric BC to verify that the discretization of the Maxwell equations does not generate spurious splitting. As can be expected, no differences were observed between the two cases.

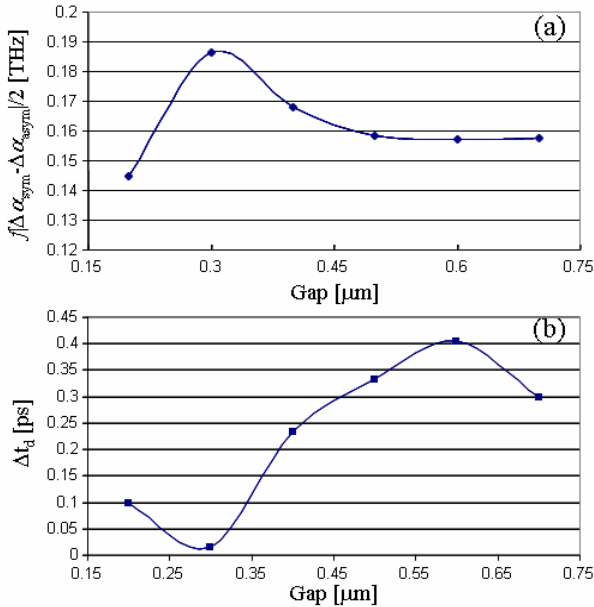

Figure 5. Dependence of the differences in the self-frequency shifts (a) and in the delay per resonator (b) on the inter-resonator gap.

Thus, the numerical analysis confirms that the symmetric and anti-symmetric Bloch wave bands are non-degenerate and, therefore, a signal which is injected to a single port of the CROW might undergo substantial distortion as it propagates in the structure.

Next, we set to investigate the impact of the parameters of the CROW ($W_{gap}$, $m$, etc.) on the band splitting and the coefficients of the dispersion relations ($\Delta\alpha$ and $\kappa$). Figure 5 depicts the dependence of differences in the self frequency shift and in the time delay per resonator, given by $t_d=1/\kappa\Omega$, for the studied structure on the inter-resonator gap. Non-negligible differences in the self frequency shift and even more importantly, in the delay per resonator are found for the different bands. For example, for a 25 resonator CROW with $W_{gap}$ of 0.5μm, the difference between the delays of the different symmetries (which are equally excited) exceeds 8ps. For a 40Gbs return-to-zero (RZ) data channel this difference *exceeds one third* of the symbol duration and is clearly unacceptable.

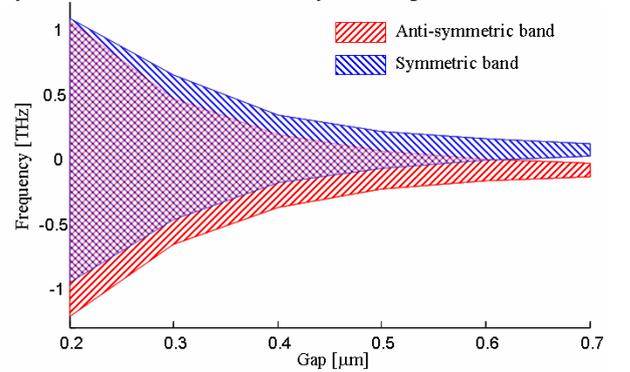

Figure 6. Transmission bands of the Anti-symmetric (red) and symmetric (blue) Bloch modes of a CROW for different gaps.

Although figure 5 indicates some of the problems induced by the symmetry properties of the CROW, it does not provide a complete view of the impact of the band non-degeneracy on the CROW performances. Figure 6 depicts the transmission bands of the symmetric and anti-symmetric branches as a function of the inter-resonator gap. When the gap is small (strong coupling) there is substantial overlap between the transmission bands. However, as the gap is increased, the overlap decreases rapidly and vanishes completely for $W_{gap}>0.6$μm. Thus, the usable bandwidth of the CROW is *reduced substantially* by the non-degeneracy of the two branches and *vanishes* completely for $W_{gap}>0.6$μm. Note that for $W_{gap}=0.6$μm the bandwidth of each branch separately is approximately 30GHz, which is still useful for telecommunications applications but there is no overlap between the bands.

Figure 7 depicts the impact of the angular modal number, $m$, on the transmission bands of the two branches for $W_{gap}=0.5$μm. As the angular modal number is increased the overlap between the transmission bands decreases although non-monotonically. This trend can be understood in view of the impact of the gap (Figure 5). Larger angular modal numbers correspond to higher frequencies and thus, tighter confinement and lower coupling coefficients. As shown in figure 5, the transmission bands overlap decreases as the coupling is reduced and, therefore, it can be expected that similar behavior will be shown for larger $m$.

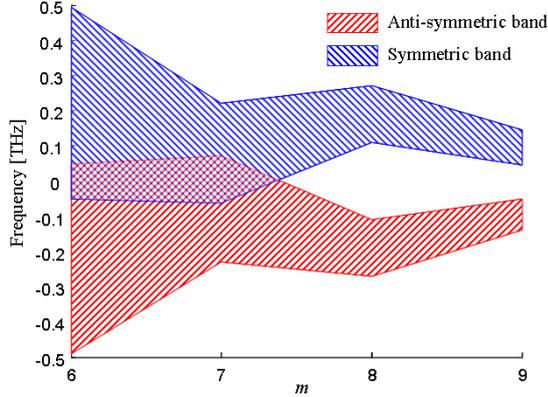

Figure 7. Transmission bands of the Anti-symmetric (red) and symmetric (blue) Bloch modes of a CROW for different $m$'s.

## V. DELAY, BANDWIDTH AND BIT STORAGE

Integrated optics slow light structures such as CROWs have been the focus of numerous studies because of their diverse potential applications in the field of optical communications, in particular – delay lines. However, the results of the structure analyzed in section IV indicate that the symmetry induced degeneracy lifting may result in substantial distortion of the transmitted signal. Thus, it is important to quantify the impact of this, additional, modal dispersion on the performances of CROW structures for optical delay line applications.

To analyze the additional modal dispersion, we follow a similar approach to the one presented by Khurgin [20]. We consider an RZ data-stream where the time slot allocated for each bit is $\Delta T_{sig}=B^{-1}$ where $B$ is the bit-rate. We assume a Gaussian pulse with its FWHM equals $\frac{1}{2}\Delta T_{sig}$. The FWHM of the signal spectrum is given by $\Delta\omega_{sig}=8\ln(2)B$. The signal is launched into one of the input ports, thus equally exciting the symmetric and anti-symmetric dispersion branches. The delay of the signal propagating through a CROW consisting of $N_r$ resonators depends of the frequency and the specific branch:

$$T_d^{S,AS} \approx \frac{N_r}{\kappa_{S,AS}\Omega} + \frac{\frac{1}{2}N_r \cdot (\omega-\Omega-\frac{1}{2}\Delta\alpha_{S,AS})^2}{(\kappa_{S,AS}\Omega)^3} \quad (12)$$

where the indices S and AS indicate respectively the symmetric and anti-symmetric bands. In order to efficiently exploit the overlapping bandwidths of the two branches it is desired to set the carrier frequency to the center of the overlap regain:

$$\Omega_0 = \Omega(1+\tfrac{1}{2}\delta\Delta\alpha) \quad (13)$$

where $\delta\Delta\alpha=\frac{1}{2}(\Delta\alpha_S+\Delta\alpha_{AS})$. Thus we can express the time delays of the different branches as:

$$T_d^{S,AS} \approx \frac{N_r}{\kappa_{S,AS}\Omega} + \frac{\frac{1}{2}N_r \cdot (\Delta\omega-I_{S,AS}\frac{1}{2}\Omega\delta\Delta\alpha)^2}{(\kappa_{S,AS}\Omega)^3} \quad (14)$$

where $I_S=1$, $I_{AS}=-1$ and $\Delta\omega=\omega-\Omega_0$. Determining the differential time delay is more involved here because it depends on the specific parameters of the dispersion relations of the two branches. In view of figures 4 and 5, we find that for the structure analyzed here, $\Delta\alpha_S>\Delta\alpha_{AS}$ and $\kappa_S<\kappa_{AS}$. Therefore, the minimal and maximal time delays in the bandwidth are given at the anti-symmetric and the symmetric branches respectively:

$$T_d^{\max} = \frac{N_r}{\kappa_S\Omega} + \frac{\frac{1}{2}N_r \cdot (\Delta\omega+\frac{1}{2}\Omega\delta\Delta\alpha)^2}{(\kappa_S\Omega)^3} \quad (15)$$

$$T_d^{\min} = \frac{N_r}{\kappa_{AS}\Omega}$$

Requiring that the maximal differential time delay does not exceed a quarter of the bit interval yields the following limitation on the bandwidth:

$$\frac{N_r(\kappa_S^{-1}-\kappa_{AS}^{-1})}{\Omega} + \frac{\frac{1}{2}N_r \cdot \left[4\ln(2)B+\frac{1}{2}\Omega\delta\Delta\alpha\right]^2}{(\kappa_S\Omega)^3} \leq \frac{1}{4B} \quad (16)$$

Figure 8 depicts the maximal delay that can be provided without distortion by the CROW discussed in section IV for a given signal bit rate $B$. The delay is defined at the average of the delays of the symmetric and anti-symmetric branches at the carrier frequency $\Omega_0$.

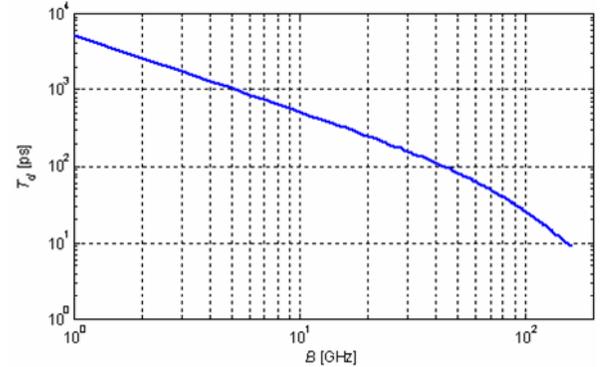

Figure 8. Maximal delay as a function of the bandwidth.

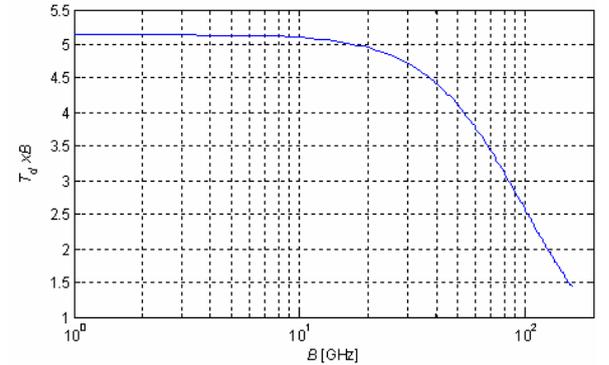

Figure 9. Delay-bandwidth product as a function of the bandwidth.

Below 30Gbs, the delay and the bandwidth are inversely proportional, indicating a relatively constant delay-bandwidth product. However, as the bit-rate exceeds 30Gbs the achievable delay decreases more rapidly indicating a decrease in delay-bandwidth product. This can also be seen in figure 9 which depicts the dependence of the delay-bandwidth product on the bit-rate. At low bit rates ($B<10Gbs$) the delay-bandwidth product of the CROW is approximately 5 but decreases rapidly reaching unity as the $B$ exceeds 200Gbs.

Another important factor in the determination of the achievable delays and bit-rates is the quality factor ($Q$) of the micro-resonators composing the CROW. As shown previously, loss could be the dominant limiting mechanism especially at low bit-rates [20]. Here, however, it is difficult to derive a simple expression for the impact of loss because, unlike the single branch CROW, the loss cannot be modeled as a simple band-pass filter. Because of the frequency shift between the symmetric and anti-symmetric branches, the loss-induced band-pass filtering is not centered on the carrier frequency and, thus, it is difficult to derive simple relation between the bandwidth and the corresponding pulse duration. In order to evaluate the impact of the microdisks $Q$, as well as additional system considerations such as noise and higher order dispersion, a comprehensive numerical analysis of the pulse propagation is required.

## VI. IMPACT ON LINK PERFORMANCES

The impact of the CROW characteristics such as dispersion, loss, etc. on the communication link performance are commonly expressed by the decrease in eye-pattern opening compared to the back-to-back case (i.e. without the CROW) [21]. The eye-pattern opening represents a measure to the available noise margin around the decision threshold employed to determine the logical value of the bit.

Similar to the previous section we assume light is launched into one of the input ports, thus exciting both dispersion branches. Each excited branch accumulates different phase and the signals at the output ports are superpositions (sum and difference) of the signals at each branch:

$$E_{out}^1 = \tfrac{1}{\sqrt{2}} E_{in}[\exp(-iK_S \Lambda N_r) + \exp(-iK_{AS} \Lambda N_r)]$$
$$E_{out}^2 = \tfrac{1}{\sqrt{2}} E_{in}[\exp(-iK_S \Lambda N_r) - \exp(-iK_{AS} \Lambda N_r)]$$
(17)

where $K_{S,AS}$ are the Bloch wavenumbers of the symmetric/anti-symmetric branches. The relevant output port must be chosen with caution as it depends on the number of rings comprising the CROW. If the two symmetry branches were degenerate (as it is commonly assumed) then for a single input port excitation, the Bloch modes interfere constructively in sum (difference) output port for an even (odd) number of rings. However, because the two branches are not degenerate, this simple rule does not hold and the more dominant output port (if there is one) is determined by the specific splitting of the symmetry branches at the carrier frequency.

To proceed, we assume the best scenario where it is possible to choose the output port with the maximal eye-pattern opening. Although, at first, such assumption may seem unrealistic, it should be kept in mind that $N_r$ is *fixed* for a given CROW and therefore, the better output port can be predetermined. Figure 10 depicts the eye-opening penalty dependence on the number of resonators (solid blue line) for a 40Gbs RZ signal propagating through the CROW analyzed in section V. The pulses in the signal are assumed to be in the standard 50% duty cycle RZ format. For comparison, the figure also shows the penalty when only one of the branches – the symmetric (green stars) or anti-symmetric (red circles) are used. As shown in the figure, the impact of the branch splitting on the link performances is substantial. While the penalty of the dispersion of the each branch is practically negligible (less than -0.1dB for a 50 resonator CROW), the penalty when both branches are excited could reach several dB (which is unacceptable) even for a small number of cavities.

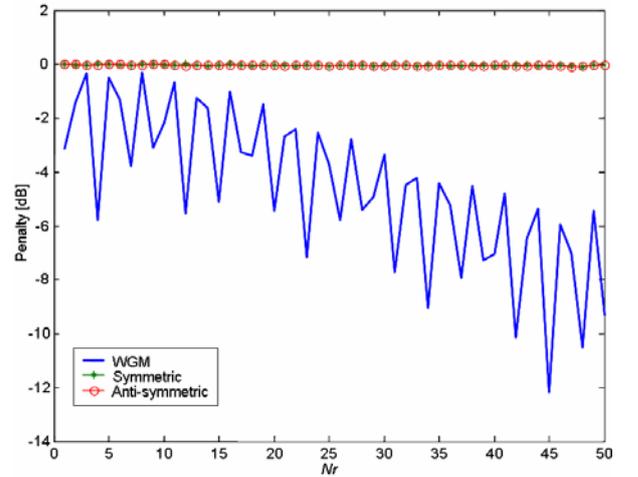

Figure 10. Eye-pattern opening penalty as a function of the number of resonators for a 40Gbs RZ signal.

Figure 11 depicts the penalty as a function of $N_r$ for different data-rates. Below 40Gbs, the link penalty rapidly varies between 0 and -4dB but without noticeable global degradation. For 40Gbs signal, however, in addition to the rapid variations there is also a clear overall increase in the eye-opening penalty.

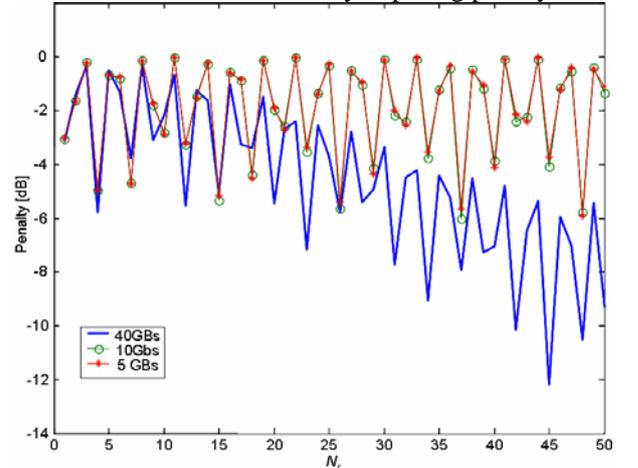

Figure 11. Eye-opening penalty as a function of the number of resonators for different data-rates.

The eye-opening penalty presented in figures 10 and 11 is caused by dispersion only. Nevertheless,

losses in the resonators are also an important limiting factor [20]. For each symmetry branch, the loss in the CROW is proportional to the time delay, i.e.:

$$P_{out}(\omega) = P_{in}(\omega)e^{-\alpha\frac{c}{n}T_d(\omega)} = P_{in}(\omega)e^{-\frac{\Omega}{Q}T_d(\omega)} \quad (18)$$

where $T_d(\omega)$ is the delay given by (14), $\alpha$ is the propagation loss coefficient and $Q$ is the quality factor of an individual resonator. The loss can be readily introduced into (17).

Figure 12 depicts the penalty of the finite $Q$ of the cavities on the eye-pattern opening for a 40 Gbs RZ signal. As long as the $Q$ exceeds $10^6$, its impact on the link performances is rather negligible. However, for $Q=10^5$ (or lower) additional penalty of approximately ~0.13dB per resonator is clearly observed.

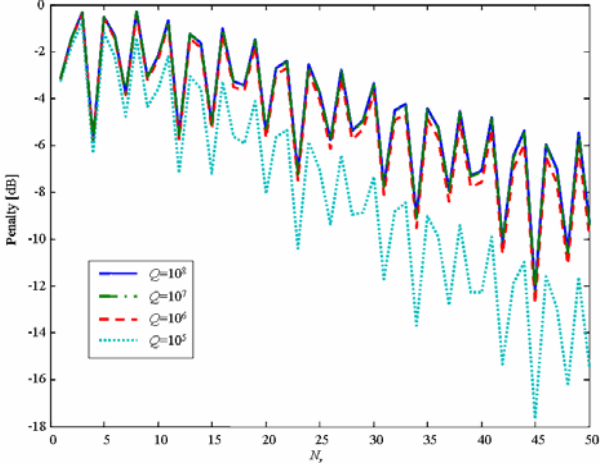

Figure 12. Eye-pattern opening penalty as a function of the number of resonators for different $Q$ factors.

For comparison, figure 13 shows the penalty of the finite $Q$ when only the symmetric CROW branch is used. Clearly, the impact of the loss is similar in both cases although additional penalty appears at $Q=10^6$ for the later case. Nonetheless, it is likely that this additional penalty exists also when both branches are excited but is masked by the rapid oscillations in Fig. 12.

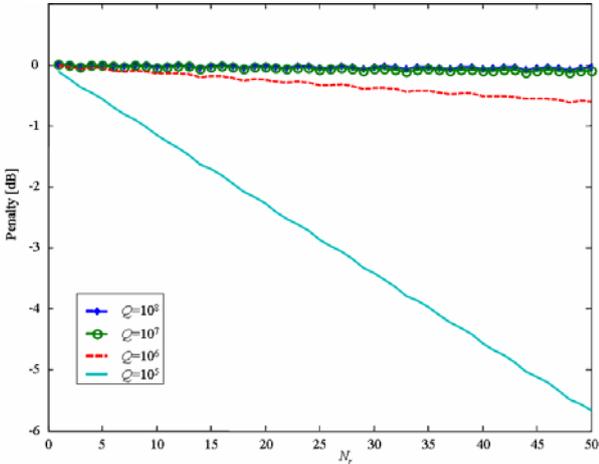

Figure 13. Eye-pattern opening penalty as a function of the number of resonators for different $Q$ factors for the symmetric branch only.

## VII. DISCUSSON AND SUMMARY

Microring or a microdisk based CROWs exhibit inherent band splitting stemming from the symmetry properties of the structure. Group theoretical analysis of the CROW structures shows that degeneracy of the CW and CCW propagating waves of the individual unit cells is lifted. Consequently, two separate passbands corresponding to the symmetric and anti-symmetric Bloch modes are formed.

The splitting of the bands can be understood intuitively by considering the dependence of the overlap integrals comprising $\Delta\alpha$ and $\kappa$ in the modal field symmetry [19]. Consider, for example, the $m=1$ angular modal number (see Fig. 14). It is clear that the overlap integrals between the field profiles (determining $\kappa$) and between the mode profile and the adjacent resonators index profile (determining $\Delta\alpha$) of the symmetric and anti-symmetric modes differ. Consequently, the coupling coefficients and self frequency shifts of the two branches are not identical – in contrast to naive intuition.

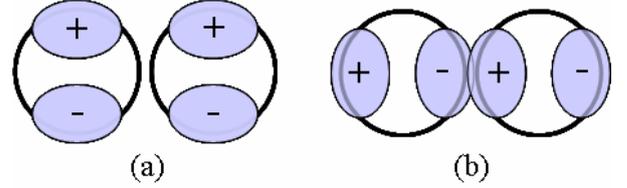

Figure 14. Schematic modal field profiles of the (a) anti-symmetric and (b) symmetric modes for $m=1$.

The differences between the coupling coefficients and, particularly, the self-frequency shift could be substantial – resulting in completely non-overlapping passbands. Larger inter-resonator gaps and higher angular modal numbers seem to decrease the spectral overlap differences between the branches. This can be understood in view of figure 14: larger gaps reduce the coupling coefficients and narrow the passbands while maintaining similar self-frequency shifts [see Fig. 5(a)]. Higher angular modal numbers correspond to higher frequency and stronger radial confinement of the field in the individual resonators, hence generating a similar effect.

We also studied the impact of the symmetry branch splitting on the achievable performances of a CROW delay line and the additional penalty of the effect when such CROW is incorporated into a communication link. We found that the additional dispersion induced by symmetry branch splitting induces substantial penalty which significantly limits the number of resonators and, hence, the achievable delay. The additional penalty as a function of $N_r$ is manifested in two forms: 1) a constant decrease of the eye-pattern opening and 2) rapid variation in the order of several dBs caused by the overall phase difference between the two branches. The later is of supreme importance as it may generate unacceptable penalty (more than 2dB) even for small number of resonators.

On the other hand, when data is transmitted using one of the natural Bloch modes of the CROW (either the symmetric or the anti-symmetric), both forms of the excessive penalty are significantly reduced. In particular, the rapid variations vanish completely, thus enabling the employment of long CROWs and the generation of long delays.

In addition, we studied the impact of propagation losses in the individual cavities (or their *Q*) on the communication link. As can be expected, once the *Q* drops below a threshold value ($10^6$ in our case), the propagation losses becomes an important factor limiting the length of the CROW and the achievable delay. However, we have not observed any significant improvement (*as far as loss is concerned*) when one of the CROW Bloch modes is used for the data transmission. Therefore, it seems that the primary impact of the symmetry branches splitting is manifested by the additional dispersion and de-phasing it induces.

To summarize, we present a new dispersion mechanism in CROW delay lines. The additional dispersion stems from the symmetry properties of the structure and may result in substantial limitation of the achievable delay and overall performance. Note, that our analysis assumes complete degeneracy of the modes of the individual resonators. Thus, the impact of mode splitting caused by surface roughness is neglected. However, it has been shown that the roughness induced mode splitting in high-*Q* cavities is in the order of tens of MHz [17], while the splitting discussed here is of the order of several hundreds of GHz, i.e. 4 orders of magnitudes larger.

An immediate consequence of the results shown here is that the practical realization of ultra-compact, coupled cavity based, integrated optical delay lines requires much care. In particular, it is highly desired to excite one of the natural Bloch modes of the CROW and not the WGM-like solutions in order to avoid the additional dispersion induced by the symmetry branch splitting. This could be readily achieved, e.g., by splitting the incoming signal using a Y-branch and launching light into both import ports (port 1 and port 2 in Fig. 1). This arrangement excites only the symmetric branch of the CROW and avoids the additional dispersion. Finally, it should be emphasizes, that although the analysis shown here is restricted to CROW, other microring based structures such as SCISSOR [10] and SC-CROW [22] are expected to exhibit similar properties because, like the CROW, they also possess $C_{2v}$ symmetry properties.

## ACKNOWLEDGMENTS

J. S. thanks Jacob Khurgin for useful discussion regarding the analysis in Section V. The support of the Israeli Ministry of Science and Technology and the Israeli Science Foundation is gratefully acknowledged.

## APPENDIX: PROPERTIES OF THE $C_{2v}$ SYMMETRY GROUP

The $C_{2v}$ group symmetry includes four covering operations – Identity (E), 180° rotation around the *z* axis ($C_2$), *x*-axis mirror inversion ($\sigma_x$), and *y*-axis mirror inversion ($\sigma_y$). Using the 4-dimensioal vector representation (1), the group operations can be represented by 4×4 matrices:

$$E = \begin{pmatrix} 1 & 0 & 0 & 0 \\ 0 & 1 & 0 & 0 \\ 0 & 0 & 1 & 0 \\ 0 & 0 & 0 & 1 \end{pmatrix}; \quad C_2 = \begin{pmatrix} 0 & 0 & 1 & 0 \\ 0 & 0 & 0 & 1 \\ 1 & 0 & 0 & 0 \\ 0 & 1 & 0 & 0 \end{pmatrix} \quad (A1)$$

$$\sigma_x = \begin{pmatrix} 0 & 1 & 0 & 0 \\ 1 & 0 & 0 & 0 \\ 0 & 0 & 0 & 1 \\ 0 & 0 & 1 & 0 \end{pmatrix}, \quad \sigma_y = \begin{pmatrix} 0 & 0 & 0 & 1 \\ 0 & 0 & 1 & 0 \\ 0 & 1 & 0 & 0 \\ 1 & 0 & 0 & 0 \end{pmatrix}$$

Each operation belongs to its own class and the character table of the group is given by:

| $C_{2v}$ | E | $C_2$ | $\sigma_x$ | $\sigma_y$ |
|---|---|---|---|---|
| $A_1$ | 1 | 1 | 1 | 1 |
| $A_2$ | 1 | 1 | -1 | -1 |
| $B_1$ | 1 | -1 | 1 | -1 |
| $B_2$ | 1 | -1 | -1 | 1 |

Table A1 – Character table of the $C_{2v}$ group